\pdfoutput=1
\documentclass[aps,pra,superscriptaddress,floatfix,10pt,twocolumn]{revtex4-1}
\usepackage[utf8]{inputenc}
\usepackage{graphicx,graphics}
\usepackage{dcolumn}
\usepackage{amsmath,amssymb,amsfonts}
\usepackage{amsthm}
\usepackage{latexsym,verbatim}
\usepackage{bm,dsfont}
\usepackage{ulem}
\usepackage{dsfont}
\usepackage{overpic}
\usepackage[breaklinks=true,colorlinks,citecolor=blue,linkcolor=blue,urlcolor=blue]{hyperref}
\usepackage{array}

\usepackage{float}
\usepackage{nicefrac}
\usepackage{overpic}
\usepackage{pifont}
\usepackage{tikz}
\usepackage{quantikz}
\newcommand{\cmark}{\ding{51}}
\newcommand{\xmark}{\ding{55}}
\newtheorem{theorem}{Theorem}
\newtheorem{corollary}[theorem]{Corollary}

\usepackage[most]{tcolorbox} 
\newtcolorbox{theorybox}{
    colback=black!5,  
    colframe=black,     
    breakable,          
    enhanced,        
    width=\textwidth    
}

\footnotetext{Equal contribution.}

\begin{document}

\title{
Building globally controlled quantum processors with ZZ interactions 
}
\author{Roberto Menta$^\star$}
\email{rmenta@planckian.co}
\affiliation{Planckian srl, I-56127 Pisa, Italy}
\affiliation{NEST, Scuola Normale Superiore, I-56127 Pisa, Italy}
\author{Francesco Cioni$^\star$}
\email{francesco.cioni@sns.it}
\affiliation{NEST, Scuola Normale Superiore, I-56127 Pisa, Italy}
\author{Riccardo Aiudi}
\affiliation{Planckian srl, I-56127 Pisa, Italy}
\author{Francesco Caravelli}
\affiliation{Planckian srl, I-56127 Pisa, Italy}
\affiliation{Dipartimento di Fisica dell’Universit\`{a} di Pisa, Largo Bruno Pontecorvo 3, I-56127 Pisa, Italy}
%
\author{Marco Polini}
\affiliation{Planckian srl, I-56127 Pisa, Italy}
\affiliation{Dipartimento di Fisica dell’Universit\`{a} di Pisa, Largo Bruno Pontecorvo 3, I-56127 Pisa, Italy}
\author{Vittorio Giovannetti}
\affiliation{Planckian srl, I-56127 Pisa, Italy}
\affiliation{NEST, Scuola Normale Superiore, I-56127 Pisa, Italy}

\begin{abstract}
We present a comprehensive framework for constructing various architectures of globally driven quantum computers, with a focus on superconducting qubits. Our approach leverages static inhomogeneities in the Rabi frequencies of qubits controlled by a 
common classical pulse -- a technique we refer to as the “crossed-qubit” method.
We detail the essential components and design principles required to realize such systems, highlighting how global control can be harnessed to perform local operations, enabling universal quantum computation. 
This framework offers a scalable pathway toward quantum processors by striking a balance between wiring complexity and computational efficiency, with potential applications in addressing current challenges to scalability.
\end{abstract}

\maketitle

\section{Introduction}
Modern gate-based quantum computing architectures rely heavily on the local controllability of qubits, with individually addressed microwave or laser pulses enabling the realization of arbitrary gate sequences. However, scaling up such architectures introduces substantial hardware overhead and calibration complexity, especially in systems like superconducting qubits, where individual control lines and crosstalk become critical bottlenecks~\cite{Kjaergaard_2020, Mohseni2024}. An alternative paradigm is to employ global control fields that act simultaneously on all -- or subsets of -- qubits, reducing hardware demands and potentially enhancing scalability.

The idea of global control in quantum computation has its roots in early theoretical proposals.
In 1993, Seth Lloyd introduced one of the first schemes for quantum computation under global control~\cite{Lloyd_1993, Lloyd_1993_SI}. In his one-dimensional architecture, qubits are grouped into three distinct species, $\{A, B, C\}$, arranged periodically in space as $ABCABC$, each species being driven by a different global control pulse. Lloyd demonstrated that, by applying sequences of globally acting yet species-selective and temporally asymmetric operations, any quantum algorithm can be implemented, thereby achieving universal quantum computation without the need for full individual qubit addressing. Shortly thereafter, Simon Benjamin proposed a simplified model requiring only two species of qubits, $A$ and $B$, arranged in an alternating pattern along a one-dimensional chain~\cite{benjamin_2000}. In this architecture, logical qubits are encoded within patches of the array and separated by buffer regions, with their mobility and manipulation achieved through species-specific global pulses. As in Lloyd's proposal, this model relies on asymmetry for computational universality: in this case, a single localized control unit (CU) traverses the array, serving as a dynamic pointer that enables conditional gates on logical qubits by virtue of its spatial overlap with them.
Benjamin’s framework spurred a series of variations and improvements in the early 2000s, many of which explored alternative control mechanisms -- ranging from modulating interaction strengths to tuning local energy gaps -- and alternative encodings of quantum information~\cite{benjamin_2001_1, benjamin_2001_2, benjamin_2003, benjamin-bose_2004}. Additional proposals generalized these ideas to other interaction types, geometries, and logical operations~\cite{Levy_2002, Raussendorf_2005, Ivanyos_2005, Ivanyos_2006, Kay_2004, Fitzsimons_2006, Fitzsimons_2007_2, Silva_2009}. Nevertheless, these models remained mostly theoretical and hardware-independent, and failed to reach the experimental maturity required to compete with architectures based on full local control.

Despite this historical stagnation, the idea of global control offers compelling advantages that have motivated renewed attention in recent years~\cite{Maslov2021, vandeWetering2021, Hu2025}. Chief among them is the potential for drastic reduction in the complexity of classical control infrastructure: eliminating the need for individual qubit lines simplifies the wiring and routing layers of the processor, a crucial bottleneck in current superconducting and solid-state platforms. This issue, known as the “wiring problem", arises from the need for multiple control signals for each qubit, leading to wiring congestion, particularly in superconducting platforms~\cite{Girvin2008, Gambetta2017, tsai2020pseudo, tsai2021gate}.
By avoiding dense wiring and minimizing the proximity of classical electronics to quantum hardware, global control can mitigate thermal load, reduce classical-quantum cross-talk, and enable denser qubit layouts. Moreover, global control schemes are naturally compatible with massive parallelism. Since control pulses act identically on entire species or spatial regions, they can implement collective operations in a single time step~\cite{Maslov2021, vandeWetering2021, Singh2025}. This intrinsic parallelism can be particularly beneficial for quantum error correction protocols that require synchronized multi-qubit gates, syndrome extraction, and repetitive stabilizer measurements~\cite{Bririd_2004, Kay_2005, Kay_2007, Fitzsimons_2007, Fitzsimons_2008, Aharonov_1996, Aharonov_2008}.
Importantly, these models also offer a potential pathway toward scalability in the post-NISQ era~\cite{Preskill-NISQ}. While current quantum processors prioritize full controllability at small scale, future large-scale architectures will face increasingly severe overheads associated with wiring, calibration, and thermal management~\cite{Kjaergaard_2020, Martinis2019quantum, Krinner2022, Mohseni2024}. Despite recent proposals to reduce number of control lines~\cite{Koch2023, Koch2023_2, Koch2025}, {\it global control} offers an elegant solution by offloading much of this overhead to global signals and exploiting architecture-level symmetry.

Recent developments have revitalized this line of research by anchoring it in experimentally relevant platforms. A universal control scheme for Rydberg atom arrays has been proposed~\cite{cesa2023universal}, where globally applied laser fields achieve local selectivity via interaction blockade. More recently, the global control paradigm has been revisited in superconducting qubits, with novel approaches that combine global microwave drives and engineered architectural asymmetries~\cite{menta2024globally, cioni2024conveyorbelt}. 

These advances have made global control experimentally viable within one of the most mature quantum computing platforms.

In this work, we generalize the “crossed-qubit” method introduced in Refs.~\cite{menta2024globally, cioni2024conveyorbelt}, further reducing the number of control lines needed to achieve universal quantum operations. 
Specifically, we show that by introducing static inhomogeneities in the Rabi frequencies of qubits driven by a common classical pulse, effective local control can be achieved.
This enables us to reduce the number of control lines from three to two in the {\it ladder} architecture of Ref.~\cite{menta2024globally}, eliminating the need for a dedicated initialization control. Similarly, we present a version of the {\it conveyor-belt} model of 
Ref.~\cite{cioni2024conveyorbelt} with two control lines, which in its original form  required three and more physical qubits. 

The paper is organized as follows. In Sec.~\ref{sec:themodel}, we introduce the notation and briefly review the physics of  the models discussed in Refs.~\cite{menta2024globally, cioni2024conveyorbelt}.
In Sec.~\ref{sec:global-vs-local}, we present a generalization of the crossed-qubit method and demonstrate how it enables
 local control over individual qubits using global pulses. 
Sec.~\ref{sec:examples} illustrates two architectural variants inspired by the schemes proposed in Refs.~\cite{menta2024globally, cioni2024conveyorbelt}. Concluding remarks and final discussions are provided in Sec.~\ref{sec:conclusion}.

\section{Notation and Preliminaries} \label{sec:themodel}

\begin{figure}[t]
\centering
\includegraphics[width=0.8\columnwidth]{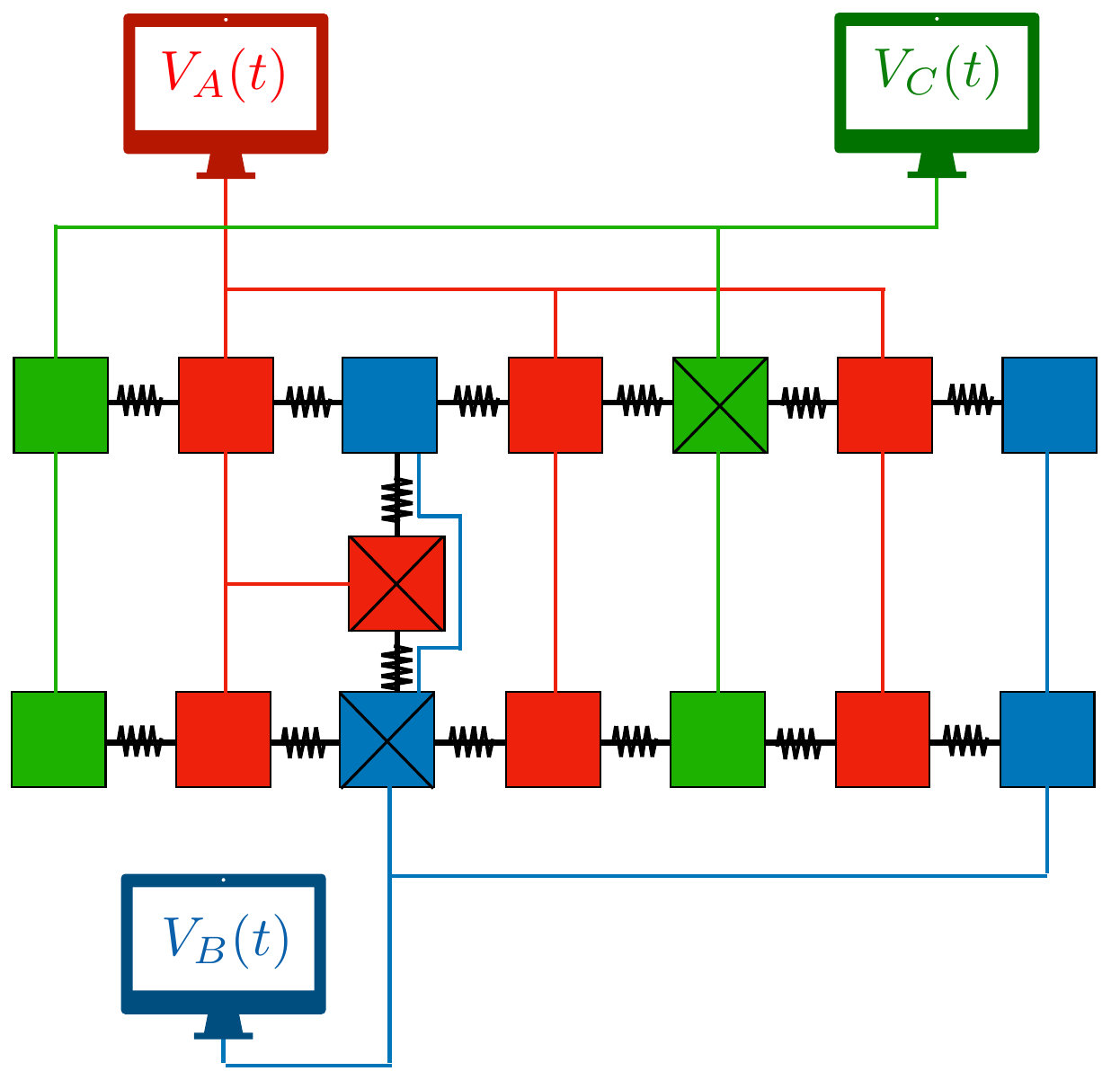}
\caption{Example of the Hamiltonian model  studied in this paper  which involves a set  of
  three  different species  of superconducting qubits, ${\cal S} =\{ A, B,C\}$, represented by the red, blue and green squares, respectively. These qubits are arranged on a graph in the form of a ladder, consisting of two rows connected by an edge.
  Black springs between pairs of qubits indicate always-on ZZ interactions. 
  Each species is globally controlled by independent classical electrical pulses delivered through dedicated wiring, shown as continuous red, blue, and green lines connecting all qubits of the corresponding group. Elements with crosses correspond to crossed-qubits.
 }
\label{fig:example}
\end{figure}

Here, we give a brief introduction to the physics of 
the models presented in  Refs.~\cite{menta2024globally, cioni2024conveyorbelt}.

\subsection{Hamiltonian}
In the globally driven superconducting architectures of Refs.~\cite{menta2024globally, cioni2024conveyorbelt},
a collection of qubits is arranged into a planar graph, where edges denote always-on nearest-neighbor ZZ interactions of uniform strength~$\zeta$.
The qubits are further partitioned into a set ${\cal S}$ of independent species, labeled by an index $\chi$, such that no two adjacent qubits belong to the same species, see Fig.~\ref{fig:example}.
This partitioning satisfies the following key requirement:
\begin{itemize}
\item[(P1)] it ensures that when constructing the planar graph, all qubits have nearest neighbors belonging to different species; 
\item[(P2)] each species corresponds
to a sub-sets of qubits that are globally controlled by the same source signal 
$V_{\chi}(t):= {\cal A}_\chi(t)\sin(\omega_{\mathrm{d},\chi}t + \phi_{\chi}(t))$
 generated by a dedicated electrical pulse source. This signal is
characterized by a constant oscillation frequency  $\omega_{\mathrm{d},\chi}$ and (potentially) time-dependent
phase $\phi_\chi(t)$  and amplitude ${\cal A}_{\chi}(t)$. 
\end{itemize} 
As a result, the number of independent control signals matches the number of species, and by design, no direct interactions occur between qubits of the same species. 


We begin with some necessary definitions. Let $\hat{\sigma}_i^{(x, y, z)}$ denote the Pauli operators of the $i$-th qubit defined 
in the local energy basis ${\vert g_i \rangle := (0,1)^{\rm T}, \vert e_i \rangle} := (1,0)^{\rm T}$. The total
Hamiltonian of the system reads 
\begin{eqnarray} \label{defHT} \hat{H}(t): = \hat{H}_0 + \hat{H}_{\rm drive}(t),\end{eqnarray}  with the static component accounting for both the nearest-neighbor ZZ couplings and the local qubit  energies $\omega_i$,
\begin{equation}\label{H0}
\hat{H}_0 :=\sum_{\chi \in \mathcal{S}}   \sum_{i\in \chi} \frac{\hbar \omega_i}{2} \hat{\sigma}^{(z)}_{i}
+  \sum_{\langle i,j \rangle} \frac{\hbar \zeta}{2} \hat{\sigma}^{(z)}_{i} \otimes \hat{\sigma}^{(z)}_{j}\;, 
\end{equation}
and the driving term describing the action of the control fields is given by

\begin{eqnarray}
\label{Hdrive}
\hat{H}_{\rm drive}(t) &=&\sum_{\chi \in \mathcal{S}} V_\chi(t) \sum_{i \in \chi}  {\cal V}_i  \hat{\sigma}^{(y)}_i\;, 
 \\
&=& \nonumber
\sum_{\chi \in \mathcal{S}} \sum_{i \in \chi} \hbar \Omega_{\chi,i}(t)\sin(\omega_{\mathrm{d},\chi}t + \phi_{\chi}(t)) \hat{\sigma}^{(y)}_i\;.
\end{eqnarray} 
Above,  ${\cal V}_i$ denotes the  coupling constant mediating the interaction of the $i$-th qubit of the  species $\chi$ with its associated classical drive
$V_{\chi}(t)$, and  
\begin{eqnarray} \Omega_{\chi,i}(t):= {\cal A}_{\chi}(t)  {\cal V}_i/\hbar\;, 
\label{rabi}
\end{eqnarray}  is the corresponding
Rabi frequency. 

For each species $\chi$, 
the local frequencies $\omega_i$
 are detuned from 
 the oscillation frequency  
$\omega_{\mathrm{d},\chi}$   by an amount proportional to the interaction strength parameter~$\zeta$:
\begin{eqnarray}
\label{coordination}
\omega_i = \omega_{\mathrm{d},\chi} - \kappa_i \zeta \;, 
\end{eqnarray} 
where $\kappa_i$ is the coordination number of the site   $i$, i.e., the number of nearest-neighbors interacting with qubit~$i$ \cite{shift}.
At this point, we exploit the fact that the  coupling constants ${\cal V}_i$ in Eq.~(\ref{rabi}) can be set during fabrication~\cite{menta2024globally}, 
and obtain that each species $\chi$ splits into two disjoint sub-groups: the set  $\chi^{\rm r}$ containing  {\it regular} $\chi$-type qubits characterized by  a  
 reference value $\bar{\cal V}$ of coupling constant, and the set $\chi^{\times}$ 
of {\it crossed} $\chi$-type qubits. In this case, the coupling constants are equal to twice the value of the regular $\chi$-type qubits. This design choice, which we refer to as ``crossed-qubit" method, 
partially breaks the global character of the control fields, ensuring that, irrespective of the choice of the pulse ${V}_\chi(t)$,  crossed qubits experience twice the Rabi frequency of the regular qubits, i.e.
\begin{eqnarray}\label{crossedtrick} 
 \left\{ \begin{array}{lll}
 i\in \chi^{\rm r} &\Longrightarrow& \Omega_{\chi,i}(t)= \Omega_{\chi^{\rm r}}(t):=
 {\cal A}_{\chi}(t)  \bar{\cal V}/\hbar \;, \\ \\
 i\in \chi^{\times} &\Longrightarrow& \Omega_{\chi,i}(t)= \Omega_{\chi^{\times}}(t):=
2  \Omega_{\chi^{\rm r}}(t)\;.
\end{array}  \right.  
\end{eqnarray} 
Notice that, in the following, we will also refer to {\it double-crossed} qubits as those with $\Omega_{\chi^{\mathbb{X}}}= 4 \Omega_{\chi^{\rm r}}(t)$.
Apart from this tuning, which is set at fabrication, the parameters ${\cal A}_\chi(t)$, $\phi_\chi(t)$, and $\omega_{\mathrm{d},\chi}$ remain independent of the index $i$, indicating that they control all qubits of type $\chi$ globally. The crossed qubit method allows for certain qubits to be addressed \textit{locally} despite the global control.

 \subsection{Dynamics} 
 The universality of the architectures of Refs.~\cite{menta2024globally, cioni2024conveyorbelt, cesa2023universal} for quantum computing operations relies critically on the proper concatenation of ordered sequences of unitary transformations of the form:
 
\begin{eqnarray} \label{defseq} \hat{U}_{\rm seq}:= \hat{U}_{\chi_n} \cdots \hat{U}_{\chi_2} \hat{U}_{\chi_1}\;,\end{eqnarray}

where each $\hat{U}_{\chi_j}$ targets a specific species during a dedicated time interval through appropriately tailored control-unitary gates. By preparing the system in specially encoded initial states that confine information on specific subsets of qubits, such sequences can be employed to move logical information into positions where it can be effectively manipulated (still using sequences of the form~(\ref{defseq})) thanks to carefully placed inhomogeneities in the form of crossed elements~\cite{menta2024globally, cioni2024conveyorbelt}, or superatoms in the Rydberg model proposed in~\cite{cesa2023universal}. Insights into how the individual transformations $\hat{U}_{\chi}$ are generated by the Hamiltonian~(\ref{defHT}) can be gained by moving to a rotating frame, and subsequently applying the Rotating Wave Approximation~(RWA).
Specifically, we apply the unitary transformation $\hat{U}_{\rm rf}(t) := \bigotimes_i e^{i\hat{\sigma}_i^{(z)} \omega_{\mathrm{d,}i}t/2}$, and then we neglect fast-oscillating  terms  at frequencies $\omega_i + \omega_{\mathrm{d},\chi}$. After these transformations, the Hamiltonian $\hat{H}(t)$  reduces to~\cite{menta2024globally}

\begin{eqnarray}\label{Hfinal}
\hat{H}_{\text{rf}}(t)&\simeq& \sum_{\chi \in \mathcal{S}} \sum_{i \in \chi} \frac{\hbar \Omega_{\chi,i}(t)}{2} \Big[e^{i\phi_{\chi}(t)}\vert g_i \rangle \langle e_i \vert + {\rm h.c.} \Big]  \nonumber \\
&&+  \sum_{\langle i,j \rangle} 2 \hbar \zeta \vert e_i e_{j} \rangle \langle e_i e_{j} \vert~,
\end{eqnarray}

where the $\Omega_{\chi,i}(t)$ take uniform values within  the subsets $\chi^{\rm r}$  and $\chi^{\times}$ according to Eq.~(\ref{crossedtrick}). 
A crucial operational requirement is to work in the strong coupling limit 

\begin{eqnarray}\label{blockaderegime} 
\eta_{\rm BR}:=|\zeta/ \Omega_{\chi^{\rm r}}|\gg 1\;. \end{eqnarray} 

As demonstrated in Ref.~\cite{menta2024globally},
 this regime leads to a blockade effect that is analogous to the one observed in Rydberg atoms~\cite{cesa2023universal}. 
Specifically, consider the scenario where, during a finite time interval ${\cal T}$, only the $j$-th qubit's driving is active. Then one can show that under the condition
(\ref{blockaderegime}), if this qubit has at least one nearest neighbor initialized in the excited state, then the control field is unable to induce any rotation on the $j$-th qubit.
In this regime, $\hat{H}_{\text{rf}}(t)$ leads to the control-unitary operator
\begin{eqnarray}\label{def_W}
\hat{W}_j(\theta_j,\bm{n}_j):= \hat{\openone}_j \otimes \hat{Q}_{\langle j \rangle}+
\hat{\mathbb{R}}_j(\theta_j,\bm{n}_j) \otimes \hat{P}_{\langle j \rangle}  \;, 
\end{eqnarray}
where $\hat{\openone}_j$ is the identity operator acting on the qubit $j$. The projectors $\hat{P}_{\langle j \rangle}$ and $\hat{Q}_{\langle j \rangle} = \hat{\openone}_{\langle j \rangle} - \hat{P}_{\langle j \rangle}$ enforce the blockade condition: $\hat{P}_{\langle j \rangle}$ projects onto  the subspace of the model where all  the nearest neighbors of the qubit $j$ are in the ground state
(unblocked condition), while $\hat{Q}_{\langle j \rangle}$ projects onto the orthogonal subspace where at least one neighbor is excited (blocked condition).
The operator $\hat{\mathbb{R}}_j(\theta_j,\bm{n}_j)$ is the single qubit rotation, and takes form

\begin{eqnarray}
\label{identityR} 
 &&\hat{\mathbb{R}}_j(\theta_j,\bm{n}_j):= \exp\left[ -i ({\theta_j}/{2})\bm{n}_j \cdot \vec{\sigma}_j\right] \\ \nonumber 
 &&\quad =\overset{\longleftarrow}{\exp}\left(-\frac{i}{2} \int_{\cal T} dt \Omega_j(t)[e^{i\phi_j(t)}\vert g_j \rangle \langle e_j \vert + {\rm h.c.}]\right)\;,  
 \end{eqnarray}
where $\overset{\longleftarrow}{\exp}[\cdots]$ denotes time-ordering, $\vec{\sigma}_j:=(\hat{\sigma}_j^{(x)}, \hat{\sigma}_j^{(y)},\hat{\sigma}_j^{(z)})$ is the vector of Pauli operators, $\theta_j\in[0,2\pi]$, and $\bm{n}_j$  is a 3D real unit vector parametrizing the rotation, determined by the control functions ${\Omega_j(t)}$ and $\phi_j(t)$. Note that in this notation, we have $\hat{\mathbb{R}}_j(2\pi,\bm{n}_j)=-\hat{\mathds{1}}_j$. We now use the fact that in the proposed architecture, no qubit has neighbors of the same species (P1), and that the controls act jointly on all the elements of the same species (P2). Therefore,   when selectively activating the control  $V_\chi(t)$  on the interval ${\cal T}$ and turning off all other controls, the evolution   induced by $\hat{H}_{\text{rf}}(t)$ applies the same transformation~(\ref{def_W})
on all elements of species $\chi$. Furthermore, due to the constraint~(\ref{crossedtrick}),  the parameters $\theta_j$, $\bm{n}_j$ assume uniform values within both the regular and the crossed subsets of $\chi$. The resulting (global) operation then takes the simpler form  

\begin{eqnarray}
\hat{U}_\chi&:=& \left.\overset{\longleftarrow}{\exp}\left(-\frac{i}{\hbar} \int_{\cal T} dt \hat{H}_{\text{rf}}(t)\right) \right|_{\eta_{\rm BR} \gg 1} \label{pulsechi}  \\ \nonumber 
&=&
\hat{W}_{\chi^{\rm r}}(
\theta^{\rm r},\bm{n}^{\rm r})  \hat{W}_{\chi^{\times}}(
\theta^{\times},\bm{n}^{\times})
 \;,
\end{eqnarray}  

where 

\begin{eqnarray} \nonumber 
\hat{W}_{\chi^{\rm r}}(
\theta^{\rm r},\bm{n}^{\rm r}) &: =&  \prod_{i\in \chi^{\rm r}} \hat{W}_{i}(
\theta^{\rm r},\bm{n}^{\rm r}) \;, \\ 
\hat{W}_{\chi^{\times}}(
\theta^{\times},\bm{n}^{\times})
&:=&  \prod_{i\in \chi} \hat{W}_{i}(
\theta^{\times},\bm{n}^{\times})  \label{pulsechidef} 
 \;,
 \end{eqnarray}  
with 
($\theta^{\rm r}$, $\bm{n}^{\rm r}$) and  ($\theta^{\times}$, $\bm{n}^{\times}$) determined via Eq.~(\ref{identityR})
by setting $\Omega_j(t)= \Omega_{\chi^{\rm r}}(t)$, $\phi_j(t)=\phi_\chi(t)$ for $\chi^{\rm r}$ and 
  $\Omega_j(t)= \Omega_{\chi^{\times}}(t)=2\Omega_{\chi^{\rm r}}(t) $, $\phi_j(t)=\phi_\chi(t)$ for $\chi^{\times}$.
A crucial observation here is that, despite possible correlations between $(
\theta^{\rm r},\bm{n}^{\rm r})$ and $(
\theta^{\times},\bm{n}^{\times })$, 
it can be shown~\cite{menta2024globally} that by properly shaping $V_\chi(t)$, one can implement arbitrary, independent control-unitary rotations on $\chi^{\rm r}$ and $\chi^{\times}$, as if two independent control sources drove them. As a result, we are effectively doubling the number of active controls that operate on the system, i.e. 
 reducing the total number of species needed to achieve (gate) universality.

It is important to recall that, since the sets  $\chi^{\rm r}$ and $\chi^{\times}$ are disjoint  and no ZZ interaction connects elements of the $\chi$ species (see property (P1)), 
the transformations defined in Eq.~(\ref{pulsechidef}) always commute. Specifically, we have \begin{eqnarray}  \Big[\hat{W}_{\chi^{\rm r}}(
\theta^{\rm r},\bm{n}^{\rm r}) ,  \hat{W}_{\chi^{\times}}(
\theta^{\times},\bm{n}^{\times})\Big]=0\;. \label{commute} 
  \end{eqnarray}
Moreover, for fixed $\chi$ and  any subset $\xi\in \{ \chi^{\rm r},\chi^{\times}\}$,  the transformations 
  $\hat{W}_{\xi}(\theta,\bm{n})$ form a group isomorphic to $SU(2)$, with the same composition rules as the  
single-qubit unitary matrices $\hat{\mathbb{R}}_i(\theta,\bm{n})$.
  In particular, for the angles  $\theta_1,\theta_2$ and unit vectors 
  $\bm{n}_1$, $\bm{n}_2$,  the composition satisfies
   \begin{eqnarray}  
   \hat{W}_{\xi}(\theta_2,\bm{n}_2)   \hat{W}_{\xi} (\theta_1,\bm{n}_1)=
   \hat{W}_{\xi}(\theta_3,\bm{n}_3)\;, \label{compo}
  \end{eqnarray}
 where $\theta_3$ and $\bm{n}_3$ are such that: 
  $\hat{\mathbb{R}}_i(\theta_2,\bm{n}_2) \hat{\mathbb{R}}_i(\theta_1,\bm{n}_1) = \hat{\mathbb{R}}_i(\theta_3,\bm{n}_3)$.
Finally, 
the inverse of $\hat{W}_{\xi}(\theta,\bm{n})$ satisfies
\begin{equation} 
\label{inv}
\hat{W}^{-1}_{\xi}(\theta,\bm{n})=\hat{W}^{\dag}_{\xi}(\theta,\bm{n})= \hat{W}_{\xi}(-\theta,\bm{n})= \hat{W}_{\xi}(\theta,-\bm{n})\;. 
\end{equation} 
The structure of these rotations is essential for the operations of the global architectures we discuss below.

\begin{figure}[t]
\centering
\includegraphics[width=1.0\columnwidth]{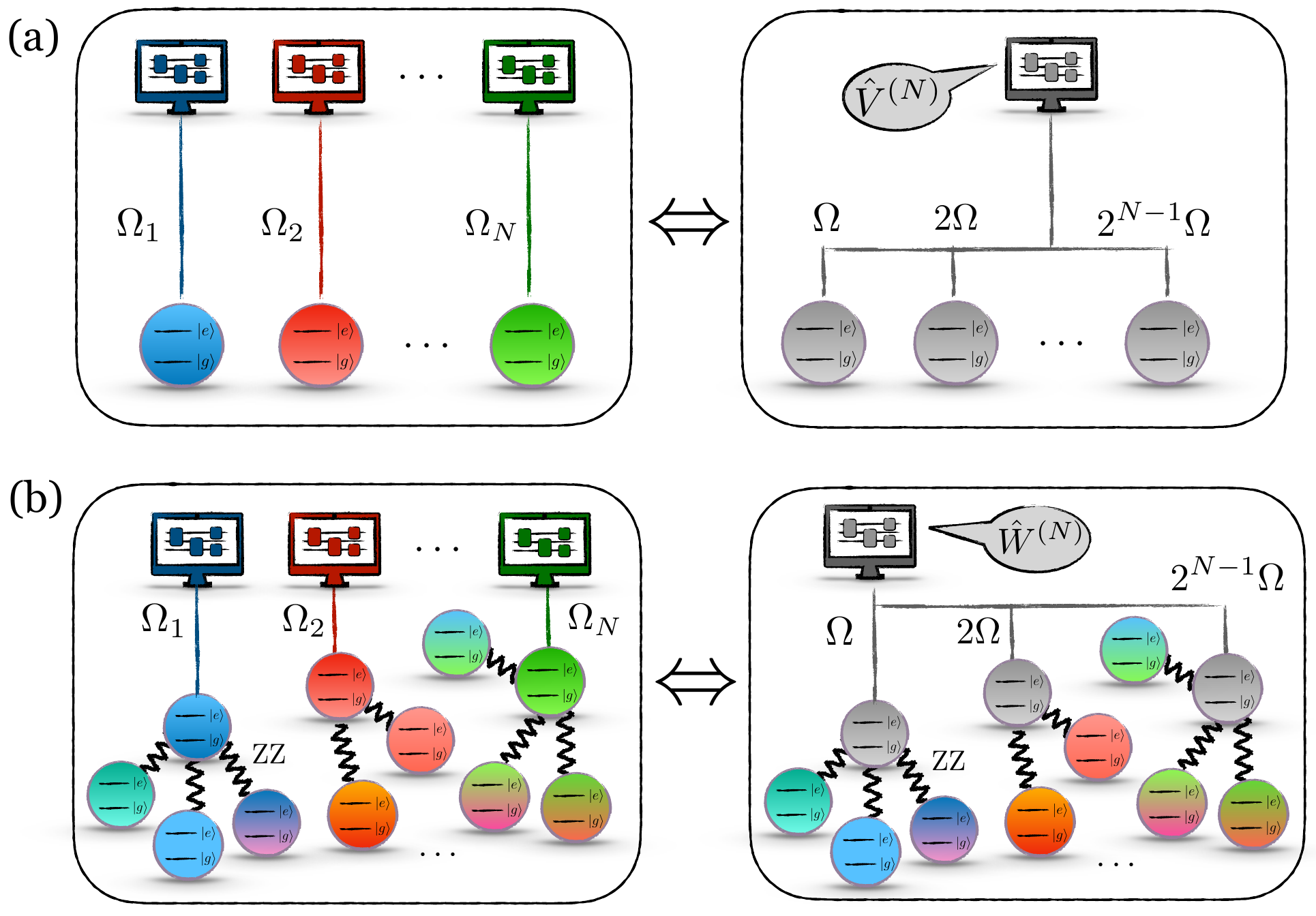}
\caption{Pictorial illustration of the equivalence between local control (left) and global control (right) for $ N $ qubits in two scenarios: non-interacting qubits [(a)] and qubits with ZZ interactions [(b)]. Here, $ \Omega_j $ (with $j = 1, \ldots, N $) denotes the Rabi frequency -- the coupling strength between the drive source and the $j$-th qubit -- associated with the control line addressing that qubit. The Rabi frequency depends on the physical properties of each qubit; thus, as shown on the left, a single drive line can induce different Rabi frequencies across the qubits. The equivalences illustrated in the figure hold under the condition specified in~\eqref{eq:rabi} (see Theorem~\ref{theorem1}). In (b), the driven qubits are coupled to other qubits of different species via ZZ interactions; in principle, these auxiliary qubits could also be addressed. The blockade regime enables the equivalence and the realization of the unitary $ \hat{W} $ stated in Corollary~\ref{corollary1}.}
\label{fig:local-global}
\end{figure}

\section{Global resources enable local control}
\label{sec:global-vs-local}

We now present a generalization of the crossed-qubit method that enables local control over multiple subsections of a given species $\chi$, while preserving the global nature of the driving pulse $V_{\chi}(t)$.
This is achieved by exploiting static non-uniformities in the Rabi couplings across the qubits.
We begin by demonstrating that arbitrary local transformations can be implemented on a set of non-interacting qubits collectively driven by a single classical field.

\subsection{Transforming global driving into local operations} \label{subsec:transf} 

Consider a collection of $N$  non-interacting  qubits,
each with uniform local energies $\hbar \omega_0$, driven by a common external classical field with amplitude
\begin{eqnarray} \label{drivingnew} V(t):= {\cal A}(t)\sin(\omega_{\mathrm{0}}t + \phi(t))\;. 
\end{eqnarray} 
Moving to the interaction picture and applying the RWA approximation, the  system's Hamiltonian takes the same form as the non-interacting part of Eq.~(\ref{Hfinal}). Specifically, we can write 
\begin{equation}
\label{Ham_rf_theorem}
    \hat{H}_{\rm int} (t) \simeq \sum_{j=1}^N \frac{\hbar \Omega_j(t)}{2} \Big [ e^{i \phi(t)} \vert g_j \rangle \langle e_j \vert + \ {\rm h.c.} \Big ] \ ,
\end{equation}
where, ${\cal V}_j$ denotes the coupling strength between the
 $j$-th qubit with the classical field,   and the corresponding Rabi frequency is  $\Omega_j(t):=  {\cal A}(t){\cal V}_j/\hbar$.
By direct integration, we find that the evolution over a time interval  ${\cal T}$ 
reduces to a product of local rotations on the individual qubits: 
\begin{eqnarray}  \label{eq1_theo}
  && \overset{\longleftarrow}{\exp}\left[-\frac{i}{\hbar} \int_{\cal T} dt \hat{H}_{\text{int}}(t)\right] \\ \nonumber
 &&\qquad = \hat{\mathbb{R}}_{1}(\theta_1,\bm{n}_1)\otimes \hat{\mathbb{R}}_{2}(\theta_2,\bm{n}_2)\otimes \cdots\otimes \hat{\mathbb{R}}_{N}(\theta_N,\bm{n}_N) \;,
\end{eqnarray}
where for each $j\in [1, \cdots, N]$, the local rotation operator $\hat{\mathbb{R}}_{j}(\theta_j,\bm{n}_j)$ is defined as in  
 Eq.~(\ref{identityR}), with  $\phi_j(t)=\phi(t)$. 
Our goal is to show that, by engineering inhomogeneous Rabi frequencies across the qubits, it is possible to design driving signals
$V(t)$ that satisfy Eq.~(\ref{eq1_theo}) for arbitrary choices of the rotation parameters $(\theta_1,\bm{n}_1)$, $\cdots$, $(\theta_N,\bm{n}_N)$, thus 
enabling full local control despite the global character of the drive, see panel (a) of Fig.~\ref{fig:local-global}. 

Thanks to the non-uniformities in the Rabi frequencies~\eqref{crossedtrick}, global unitary operations of the product form~\eqref{pulsechi} can arbitrarily rotate qubits of different species $\mathcal{S}$.  We provide a constructive proof of this statement.

We assume that the couplings follow a geometric progression:
\begin{equation}\label{eq:rabi}
    \Omega_j (t)=2^{j-1}\Omega (t) \;, \qquad \forall j\in [1,N]\;, 
\end{equation}
an assumption made for analytical convenience that can be relaxed. We consider a ``bang-bang" control scheme where both ${\cal A}(t)$ (and hence $\Omega_j (t)$) and $\phi(t)$ are piecewise constant over finite time intervals. We begin with the simpler time-invariant case:

\begin{eqnarray} \label{simplifiedchoise} 
\Omega_j (t) = \Omega_j =2^{j-1} \Omega \;, \quad \phi(t)= \phi \;, \qquad \forall t\in {\cal T}\;. 
\end{eqnarray} 

Under this condition, the time-ordering exponential in  Eq.~(\ref{eq1_theo}) 
becomes an ordinary exponential, and the rotation parameters are given by:

\begin{equation} \label{simplepara}
 (\theta_j,\bm{n}_j) = (2^{j-1}  \theta,\bm{n}_\perp)  \;, \qquad \forall j\in [1,\cdots,N]\;, 
\end{equation} 

where $\theta:=\Omega T$, with $T$ the duration of the time interval ${\cal T}$, and $\bm{n}_{\perp}: =   (\cos \phi, \sin\phi,0)$ is a 
 unit vector in the transverse (i.e., $xy$) plane. 
Accordingly, the unitary evolution induced by $\hat{H}_{\rm int} (t)$  corresponds to the operator

\begin{eqnarray}
&&\hat{V}_{1,2, \cdots, N}^{(N)}(\theta,\bm{n}_\perp)\label{pulsechi}\\  
&& \quad :=  \hat{\mathbb{R}}_{1}(\theta,\bm{n}_\perp)\otimes \hat{\mathbb{R}}_{2}(2\theta,\bm{n}_\perp)\otimes \cdots\otimes \hat{\mathbb{R}}_{N}(2^{N-1}\theta,\bm{n}_\perp)\nonumber 
 \;,
 \end{eqnarray} 

where each qubit goes through  a  rotation around the same transverse axis $\bm{n}_{\perp}$, but with angles proportional to $\theta$. 
The next step is to show that, by concatenating such operations over successive time intervals, one can implement arbitrary single-qubit rotations on all $N$ qubits, despite the global character of the drive.

\begin{widetext} 
\begin{theorybox}
\begin{theorem}[Global control enables local addressing]
\label{theorem1} Given an arbitrary set of single-qubit rotation parameters  $(\theta_1,\bm{n}_1)$, $(\theta_2,\bm{n}_2)$, $\cdots$, $(\theta_N,\bm{n}_N)$,
 there exists a finite sequence  of
angles $\theta^{(1)},\theta^{(2)}, \cdots, \theta^{(\ell)}$, and transverse unit vectors 
$\bm{n}^{(1)}_{\perp},\bm{n}^{(2)}_{\perp},\cdots, \bm{n}^{(\ell)}_{\perp}$  such that 
\begin{eqnarray}
\hat{V}_{1,2, \cdots, N}^{(N)}(\theta^{(\ell)},\bm{n}^{(\ell)}_{\perp}) \cdots
 \hat{V}_{1,2, \cdots, N}^{(N)}(\theta^{(2)},\bm{n}^{(2)}_{\perp}) \hat{V}_{1,2, \cdots, N}^{(N)}(\theta^{(1)},\bm{n}^{(1)}_{\perp})  = \hat{\mathbb{R}}_{1}(\theta_1,\bm{n}_1)
\otimes \hat{\mathbb{R}}_{2}(\theta_2,\bm{n}_2)\otimes \cdots\otimes \hat{\mathbb{R}}_{N}(\theta_N,\bm{n}_{N}) \;. \label{Tsequenzamagica}\end{eqnarray}  
\end{theorem}
\end{theorybox}
\begin{proof}
We first prove the statement for 
 $N=2$.   Let $\bm{n}_\perp$ and $\bm{m}_\perp$ be two 
 orthogonal unit vectors in the  $xy$-plane. We first recall the conjugation property of the $SU(2)$ group, which can be written as
\begin{eqnarray}
\hat{\mathbb{R}}_j(\phi, \bm{n}) \, \hat{\mathbb{R}}_j(\theta, \bm{m}) \, \hat{\mathbb{R}}_j(-\phi, \bm{n})
= \hat{\mathbb{R}}_j\left( \theta, \, {\mathcal{R}}(\phi, \bm{n}) \bm{m} \right) .
\end{eqnarray}
where ${\mathcal{R}}$ represents the rotation action on the vector. From this identity, we can see that if $ {\bm{n}}\cdot  {\bm{m}}=0$, we have ${\mathcal{R}}(\pi, \bm{n}) \bm{m}=-\bm{m}$. Then:
 \begin{eqnarray}
\hat{V}_{1,2}^{(2)}(\pi,-\bm{m}_\perp) \hat{V}_{1,2}^{(2)}(\theta/4,\bm{n}_\perp) \hat{V}_{1,2}^{(2)}(\pi,\bm{m}_\perp) \hat{V}_{1,2}^{(2)}(\theta/4,\bm{n}_\perp)  \label{eq11_theo} = \hat{\mathds{1}}_{1} \otimes \hat{\mathbb{R}}_{2}(\theta,\bm{m}_\perp)=\hat{\mathbb{R}}_{2}(\theta,\bm{m}_\perp)\;.
\end{eqnarray}
which follows from the identities: 
\begin{eqnarray}\hat{\mathbb{R}}_{j}(2\pi,\bm{n})= - \hat{\mathds{1}}_{j} \;,\qquad \qquad  
\hat{\mathbb{R}}_{j}^{-1}(\theta,\bm{m}) = \hat{\mathbb{R}}_{j}(-\theta,\bm{m}) = 
\hat{\mathbb{R}}_{j}(\theta,-\bm{m})  
= \hat{\mathbb{R}}_{j}(\pi,-\bm{n})\hat{\mathbb{R}}_{j}(\theta,\bm{m}) \hat{\mathbb{R}}_{j}(\pi,\bm{n})\; \label{eq22_theo}
\end{eqnarray}
valid for any  directions  $\bm{m}$, $\bm{n}$ which are orthogonal.
Equation~\eqref{eq11_theo} 
 implies that concatenating $\hat{V}_{1,2}^{(2)}$ operations we can 
 perform arbitrary rotations on qubit 2 around a transverse axis without affecting qubit 1. 
 Conversely, arbitrary single-qubit rotations on qubit 1 around a transverse axis can be implemented using: 
 \begin{eqnarray}
 \hat{\mathbb{R}}_{1}(\theta, \bm{n}_\perp) &=& \hat{\mathbb{R}}_{1}(\theta, \bm{n}_\perp)\otimes \hat{\mathds{1}}_{2} 
 = 
\hat{V}^{(2)}(\theta, \bm{n}_\perp) \Big(\hat{\mathds{1}}_{1} \otimes \hat{\mathbb{R}}_{2}(-2\theta, \bm{n}_\perp)\Big) \nonumber \\
&=&  \hat{V}_{1,2}^{(2)}(\theta, \bm{n}_\perp)\Big(  \hat{V}_{1,2}^{(2)}(\pi,\bm{n}_\perp) \hat{V}_{1,2}^{(2)}(\theta/2,\bm{m}_\perp) 
  \hat{V}_{1,2}^{(2)}(\pi,-\bm{n}_\perp) \hat{V}_{1,2}^{(2)}(\theta/2,\bm{m}_\perp)
\Big)\;. \label{rotation1} 
\end{eqnarray} 
We stress that the rotation axis in the above discussion is constrained to lie in the $xy$ plane. However, this is sufficient to generate arbitrary rotations in the out of plane direction as well. Indeed, Euler's theorem~\cite{Nielsen2010} states that being able to perform arbitrary rotations along two non-parallel axes allows us to generate any rotation along any desired axis. 
Therefore, by properly concatenating sequences~(\ref{eq11_theo}) and (\ref{rotation1}) we can
generate any  rotation of the form $\hat{\mathbb{R}}_{1}(\theta_1,\bm{n}_1)\otimes \hat{\mathbb{R}}_{2}(\theta_2,\bm{n}_2)$, with 
arbitrary values of $(\theta_1,\bm{n}_1)$ and $(\theta_2,\bm{n}_2)$, hence proving the thesis for $N=2$.
We now proceed to the final step of the proof by induction. 
Specifically, assuming that (\ref{Tsequenzamagica})  holds for 
$N$ qubits, 
we want to show that the same property generalizes to $N+1$ qubits.
The key idea is to extend Eq.~\eqref{eq11_theo} to higher orders using the structure of $\hat{V}^{(N)}$. 
Specifically, one can verify from
\begin{eqnarray}\label{impoide} 
\left\{ \begin{array}{l} \hat{\mathbb{R}}_{1}(\pi,-\bm{n}_\perp)\hat{\mathbb{R}}_{1}(\theta,\bm{m}_\perp) \hat{\mathbb{R}}_{1}(\pi,\bm{n}_\perp)
\hat{\mathbb{R}}_{1}(\theta,\bm{m}_\perp)
 = \hat{\mathbb{R}}_{1}(0,\bm{m}_\perp)=\hat{\mathds{1}}_{1} \;, \\\\
 \hat{\mathbb{R}}_{2}(2\pi,-\bm{n}_\perp)\hat{\mathbb{R}}_{2}(2\theta,\bm{m}_\perp) \hat{\mathbb{R}}_{2}(2\pi,\bm{n}_\perp)
\hat{\mathbb{R}}_{2}(2\theta,\bm{m}_\perp)
= \hat{\mathbb{R}}_{2}(4\theta,\bm{m}_\perp)\;,\\\\
 \hat{\mathbb{R}}_{3}(4\pi,-\bm{n}_\perp)\hat{\mathbb{R}}_{3}(4\theta,\bm{m}_\perp) \hat{\mathbb{R}}_{3}(4\pi,\bm{n}_\perp)
\hat{\mathbb{R}}_{3}(4\theta,\bm{m}_\perp)
= \hat{\mathbb{R}}_{3}(8\theta,\bm{m}_\perp)\;,\\
\\
\vdots \\\\
\hat{\mathbb{R}}_{{N+1}}(2^{N}\pi,-\bm{n}_\perp)\hat{\mathbb{R}}_{{N+1}}(2^{N}\theta,\bm{m})\hat{\mathbb{R}}_{{N+1}}(2^{N}\pi,\bm{n}_\perp)
\hat{\mathbb{R}}_{{N+1}}(2^{N}\theta,\bm{m}_\perp)
= \hat{\mathbb{R}}_{{N+1}}(2^{N+1}\theta,\bm{m}_\perp)\;.
\end{array} \right.
\label{eq2_theo}
\end{eqnarray} 
that the operation
\begin{eqnarray}\nonumber 
\hat{M}_{1,2, \cdots, N+1}(\theta,\bm{n}_\perp)&:=&  \hat{V}_{1,2, \cdots, N+1}^{(N+1)}(\pi,-\bm{m}_\perp) \hat{V}_{1,2, \cdots, N+1}^{(N+1)}(\theta/4,\bm{n}_\perp) \hat{V}_{1,2, \cdots, N+1}^{(N+1)}(\pi,\bm{m}_\perp) \hat{V}_{1,2, \cdots, N+1}^{(N+1)}(\theta/4,\bm{n}_\perp) \\
&=&\hat{\mathds{1}}_{1} \otimes 
\hat{V}_{2, \cdots, N+1}^{(N)}(\theta,\bm{n}_\perp) \;, \label{sequenzamagica}\end{eqnarray} 
emulates an $N$-qubit operation $\hat{V}^{(N)}$ on qubits 2 to $N+1$, while leaving qubit 1 untouched, using concatenations of $(N+1)$-order transformations  $\hat{V}^{(N+1)}$.
Using the induction hypothesis on the $N$-th order transformations $\hat{V}^{(N)}$, we obtain that for all arbitrary choices of  $(\theta_2,\bm{n}_2)$, $(\theta_3,\bm{n}_3)$, $\cdots$, $(\theta_{N+1},\bm{n}_{N+1})$ there exist $(\theta^{(1)},\bm{n}^{(1)}_{\perp})$, $(\theta^{(2)},\bm{n}^{(2)}_{\perp})$, $\cdots$, $(\theta^{(\ell)},\bm{n}^{(\ell)}_{\perp})$ such that 

\begin{equation}
\hat{V}_{2, \cdots, N+1}^{(N)}(\theta^{(\ell)},\bm{n}^{(\ell)}_{\perp}) \cdots
 \hat{V}_{2, \cdots, N+1}^{(N)}(\theta^{(2)},\bm{n}^{(2)}_{\perp}) \hat{V}_{2, \cdots, N+1}^{(N)}(\theta^{(1)},\bm{n}^{(1)}_{\perp})  = \hat{\mathbb{R}}_{2}(\theta_2,\bm{n}_2)
\otimes \hat{\mathbb{R}}_{3}(\theta_3,\bm{n}_3)\otimes \cdots\otimes \hat{\mathbb{R}}_{N+1}(\theta_N,\bm{n}_{N+1}) \;, \label{Ssequenzamagica}
\end{equation} 

which together with (\ref{sequenzamagica}) gives

\begin{equation} 
\hat{M}_{1,2, \cdots, N+1}(\theta^{(\ell)},\bm{n}^{(\ell)}_{\perp})\cdots \hat{M}_{1,2, \cdots, N+1}(\theta^{(2)},\bm{n}^{(2)}_{\perp})
\hat{M}_{1,2, \cdots, N+1}(\theta^{(1)},\bm{n}^{(1)}_{\perp})=\hat{\mathds{1}}_{1} 
\otimes \hat{\mathbb{R}}_{2}(\theta_2,\bm{n}_2)\otimes \cdots\otimes \hat{\mathbb{R}}_{N+1}(\theta_N,\bm{n}_{N+1}) \;. \label{Ssequenzamagica}\end{equation}  
We can also produce a sequence of $\hat{V}^{(N+1)}$'s
that induces local rotation only on the first qubit around any chosen transverse axis $\bm{n}$, invoking Euler's theorem \cite{Nielsen2010} and the identity
\begin{eqnarray}\nonumber 
 &&\hat{V}_{1,2, \cdots, N+1}^{(N+1)}(\theta,\bm{n}_\perp)\left(\hat{M}_{1,2, \cdots, N+1}(\theta^{(\ell)},\bm{n}^{(\ell)}_{\perp})\cdots \hat{M}_{1,2, \cdots, N+1}(\theta^{(2)},\bm{n}^{(2)}_{\perp})
\hat{M}_{1,2, \cdots, N+1}(\theta^{(1)},\bm{n}^{(1)}_{\perp})\right)\nonumber \\
&&\qquad \qquad \qquad = \hat{\mathbb{R}}_{1}(\theta,\bm{n}_{\perp})\otimes 
\hat{\mathds{1}}_{2} \otimes \cdots \otimes \hat{\mathds{1}}_{N+1} \;, \label{mag}\end{eqnarray}  
which holds once we take  $(\theta^{(1)},\bm{n}^{(1)}_{\perp})$, $(\theta^{(2)},\bm{n}^{(2)}_{\perp})$, $\cdots$, $(\theta^{(\ell)},\bm{n}^{(\ell)}_{\perp})$  in such a way to have 
$\theta_j = 2^{j-1} \theta$ and $\bm{n}_{j}= -\bm{n}_{\perp}$ for all $j\in [2,\cdots,N+1]$ in Eq.~(\ref{Ssequenzamagica}).
Combining Eqs.~\eqref{Ssequenzamagica} and \eqref{mag} completes the proof.
\end{proof}
\end{widetext}

It is worth noting that in Eq.~\eqref{Tsequenzamagica} the integer $\ell$
denotes the number of pulse sequences required to implement an arbitrary local operation within the globally controlled qubit set. The constructive proof presented above establishes the existence of such a
protocol, but it is not intended to provide an optimal compilation strategy. As discussed above, the recursive structure of the construction may suggest an unfavorable scaling of $\ell$ with the
number $N$ of qubit subgroups if interpreted naively. This reflects a general feature of global-control schemes: increasing $N$ inevitably increases the number of operations required to address individual subgroups. In the extreme limit where $N$ equals the total number of qubits, the protocol effectively reproduces full local control and thus loses the advantages associated with reduced wiring.
In practice, the pulse complexity depends crucially on the available control resources. The relevant ingredients include the number of control species and, for each species, the variety of accessible element types (e.g., regular, crossed, or double-crossed configurations). In the explicit architectures discussed in this work (see the next section), the use of two control species combined with crossed and double-crossed elements enables local addressing with only $\mathcal{O}(10)$ elementary operations. Moreover, the pulse sequence produced by the constructive proof should be regarded as an initial point for further optimization. In particular, many products of single-qubit rotations can be simplified using standard $\mathrm{SU}(2)$ identities, allowing for substantial reductions in the overall sequence length.

As mentioned at the beginning of the section, the result presented here does not  critically depend on the specific form of the Rabi frequencies chosen in Eq.~(\ref{eq:rabi}). Similar 
decompositions can be derived in more general scenarios where the coupling strengths satisfy 
~${\cal V}_j \neq  {\cal V}_k$ for all $j\neq\ell \in [1,N]$, although the resulting derivation is more cumbersome.  
However,  for our purposes, the precise form the inhomogeneities is not essential.  What is important  is the general principle: 
to achieve local control over individual qubits using only a single global driving field, one must break the
“global resource symmetry" by introducing physical differences between the qubits. Equation~\eqref{eq:rabi} provides a simple and effective way to do so. It is worth noting that $\mathcal{V}_j \neq \mathcal{V}_k$ ensures the absence of degeneracies in the effective control Hamiltonian, provided that the frequency separation satisfies $|\mathcal{V}_j - \mathcal{V}_k| > \delta$, where $\delta$ denotes some minimal detuning required to enable local addressing between pulses.

\subsection{Generalized crossed-qubit method} \label{subsec:gen} 
Building on Theorem~\ref{theorem1}, 
this section presents a generalization of the crossed-qubit method described in Eq.~(\ref{crossedtrick}).
The goal is to 
decompose the time evolution generated by  the driving function $V_{\chi}(t)$ into a product
of commuting,  independent, control-unitary operations, each targeting a distinct  subgroups of the species $\chi$. 
Specifically, let the species $\chi$ be partitioned into $N$ non-overlapping subsets:
 $\chi^{(1)}$, $\chi^{(2)}$, $\cdots$, $\chi^{(N)}$. For each subset $\chi^{(j)}$, we assign a uniform 
 Rabi frequency by setting 
   \begin{eqnarray}\label{genercrossedtrick} 
  i\in {\chi^{(j)}} \Longrightarrow \Omega_{\chi,i}(t)= \Omega_{\chi^{(j)}}(t):=
 {\cal A}_{\chi}(t)  {\cal V}_j/\hbar \;.   \end{eqnarray}  
Under this condition, thanks to the properties (P1) and (P2), and assuming,
as in Eq.~(\ref{pulsechi}), that only the control field  $V_{\chi}(t)$ is active on the time interval ${\cal T}$, 
 that the system operates in the blockade regime (c.f.~Eq.~(\ref{blockaderegime})) -- the time evolution generated by 
$\hat{H}_{\text{rf}}(t)$ takes the form
\begin{eqnarray}
\hat{U}_\chi&:=& \left.\overset{\longleftarrow}{\exp}\left[-\frac{i}{\hbar} \int_{t_1}^{t_2} dt \hat{H}_{\text{rf}}(t)\right] \right|_{\eta_{\rm BR} \gg 1} \label{pulsechi2}  \\ \nonumber 
&=& \hat{W}_{\chi^{(1)}}(
\theta_1,\bm{n}_1)   \hat{W}_{\chi^{(2)}}(
\theta_2,\bm{n}_2) \cdots  \hat{W}_{\chi^{(N)}}(
\theta_N,\bm{n}_N)
 \;,
 \end{eqnarray}  
 where each operator $\hat{W}_{\chi^{(j)}}(
\theta_j,\bm{n}_j)$ acts on the corresponding subset as
   \begin{eqnarray}
 \hat{W}_{\chi^{(j)}}(
\theta_j,\bm{n}_j)   &: =&  \prod_{i\in \chi^{(j)}} \hat{W}_{i}(
\theta_j,\bm{n}_j) \;,  \label{pulsechidef1} 
 \end{eqnarray}  
with $\theta_j$ and $\bm{n}_j$ defined as in 
Eq.~(\ref{identityR}). 
Notice that, due to property (P1), the operators $\hat{W}_{\chi^{(j)}}(
\theta_j,\bm{n}_j)$ and $\hat{W}_{\chi^{(j')}}(
\theta_j',\bm{n}_j')$ commute when $j\neq j'$, i.e.   
\begin{eqnarray} 
[\hat{W}_{\chi^{(j)}}(
\theta_j,\bm{n}_j),\hat{W}_{\chi^{(j')}}(
\theta_j',\bm{n}_j')] = 0 \;,
\end{eqnarray}  
just as in Eq.~(\ref{commute}).

Our goal is to show that, by  off-detuning the Rabi frequencies defined in~(\ref{genercrossedtrick}) such that 
 $\Omega_{\chi^{(j)}}(t) \neq \Omega_{\chi^{(j')}}(t)$ for all $j\neq j'$, 
 we can identify driving signals $V_{\chi}(t)$ that implement the transformation 
(\ref{pulsechidef1}) for  arbitrary choices of the parameters $(\theta_1,\bm{n}_1)$, $\cdots$, $(\theta_N,\bm{n}_N)$,
hence enabling independent control over  each $\chi^{(j)}$ using global drive --
see panel~(b) of Fig.~\ref{fig:local-global}.
The proof follows directly from the results of the previous section,  by noticing that, 
for fixed $j\in[1,\cdots,N]$, the operators 
$\hat{W}_{\chi^{(j)}}(
\theta_j,\bm{n}_j)$ obey the same algebra of the single-qubit rotations $\hat{\mathbb{R}}_{j}(\theta,\bm{n})$.
In particular 
they fulfill  the properties~(\ref{compo}) and (\ref{inv}) which allow us to rephrase the derivation of Theorem~\ref{theorem1} in terms of  the $\hat{W}_{\chi^{(j)}}(
\theta_j,\bm{n}_j)$'s. 
Specifically, let us fix the Rabi frequencies as in Eq.~(\ref{simplifiedchoise}), and focus on bang-bang control
scenarios where both 
${\cal A}(t)$ (i.e. $\Omega_j (t)$) and $\phi(t)$ are chosen to be step-continuous functions that take constant values on finite time intervals. In this case the rotation parameters of Eq.~(\ref{pulsechi2}) can be expressed as in Eq.~(\ref{simplepara}) and the transformation reduces to 
\begin{eqnarray}
    \label{eq1_theo12}
    &&\hat{W}_{\chi}^{(N)}(\theta,\bm{n}_\perp) \\ 
    \nonumber&&\;:= \hat{W}_{\chi^{(1)}}(
\theta,\bm{n}_\perp)   \hat{W}_{\chi^{(2)}}(
2\theta,\bm{n}_\perp) \cdots  \hat{W}_{\chi^{(N)}}(
2^{N-1}\theta,\bm{n}_\perp) \;.
\end{eqnarray}
Then we can state the following result:  
\begin{widetext} 
\begin{theorybox}
\begin{corollary}[Case of interacting qubits]
\label{corollary1}
Given  $(\theta_1,\bm{n}_1)$, $(\theta_2,\bm{n}_2)$, $\cdots$, $(\theta_N,\bm{n}_N)$
an arbitrary  set of $N$ rotation parameters,
 there exists a finite collection of
angles $\theta^{(1)},\theta^{(2)}, \cdots, \theta^{(\ell)}$, and transverse unit vectors 
$\bm{n}^{(1)}_{\perp},\bm{n}^{(2)}_{\perp},\cdots, \bm{n}^{(\ell)}_{\perp}$  such that 
\begin{eqnarray}
\hat{W}_{\chi}^{(N)}(\theta^{(\ell)},\bm{n}^{(\ell)}_{\perp}) \cdots
 \hat{W}_{\chi}^{(N)}(\theta^{(2)},\bm{n}^{(2)}_{\perp}) \hat{W}_{\chi}^{(N)}(\theta^{(1)},\bm{n}^{(1)}_{\perp})  = \hat{W}_{\chi^{(1)}}(
\theta_1,\bm{n}_1)   \hat{W}_{\chi^{(2)}}(
\theta_2,\bm{n}_2) \cdots  \hat{W}_{\chi^{(N)}}(
\theta_N,\bm{n}_N) \;. \label{TsequenzamagicaINT}\end{eqnarray} 
\end{corollary}
\end{theorybox}
\begin{proof}
The proof is completely analogous to that of Theorem~\ref{theorem1}, after replacing $\hat{\mathbb{R}}$ with $\hat{W}$ everywhere. 
\end{proof}
\end{widetext} 
As in the case of Theorem~\ref{theorem1}, it is worth stressing that
the above result does not depend crucially on the special choice of the Rabi frequencies as in Eq.~(\ref{simplifiedchoise}).
An application of Corollary~\ref{corollary1} will be discussed in the examples of the next section, where we will make use of a decomposition of the species involved in the model into three subsets.

\section{Examples of globally-driven quantum computers}\label{sec:examples}

In this section, we present two variants of globally driven superconducting quantum computing architectures, based on recent advances by some of the authors in Refs.~\cite{menta2024globally, cioni2024conveyorbelt}. These alternative designs are grounded in the result of Corollary~\ref{corollary1}. In particular, both schemes rely on the presence of {\it crossed} and {\it double-crossed} qubit elements. Each qubit species $\chi \in \mathcal{S} = \{A, B\}$ is subdivided into $N=3$ subtypes: regular ($\chi^{\mathrm{r}}$), crossed ($\chi^{\times}$), and double-crossed ($\chi^{\mathbb{X}}$), with Rabi frequencies defined as in Eq.~(\ref{simplifiedchoise}).
In this context, in the bang-bang scenario, crossed and double-crossed elements refer to qubits with enhanced Rabi frequencies, specifically $\Omega_{\chi^{\times}} = 2\Omega_{\chi^{\mathrm{r}}}$ and $\Omega_{\chi^{\mathbb{X}}} = 4\Omega_{\chi^{\mathrm{r}}}$. Corollary~\ref{corollary1} then implies that a generic global unitary acting on all qubits of type $\chi$ can be decomposed as:
\begin{align}
&\hat{W}_{\chi}(\theta',\bm{n}'; \theta'',\bm{n}''; \theta''',\bm{n}''') := \notag \\
&\quad \hat{W}_{\chi^{\mathrm{r}}}(\theta', \bm{n}') \, 
       \hat{W}_{\chi^{\times}}(\theta'', \bm{n}'') \, 
       \hat{W}_{\chi^{\mathbb{X}}}(\theta''', \bm{n}''') \;,
\label{double-crossed}
\end{align}
where the operators $\hat{W}_{\chi^{\mathrm{r}}}(\theta', \bm{n}')$, $\hat{W}_{\chi^{\times}}(\theta'', \bm{n}'')$, and $\hat{W}_{\chi^{\mathbb{X}}}(\theta''', \bm{n}''')$ act selectively on the regular, crossed, and double-crossed qubits, respectively.

\subsection{Ladder architecture with 2 species}

The first example of a globally driven architecture utilizing qubits of multiple Rabi frequencies is the 2D ladder structure, introduced in Ref.~\cite{menta2024globally}, comprising three species of qubits organized in a periodic pattern. 
In this section, we show a two-species variant with the same number of required physical (buffer) qubits, i.e., $\mathcal{O}(n^2)$ for $n$ computational qubits.

\begin{figure}[t]
\centering
\includegraphics[width=1.0\columnwidth]{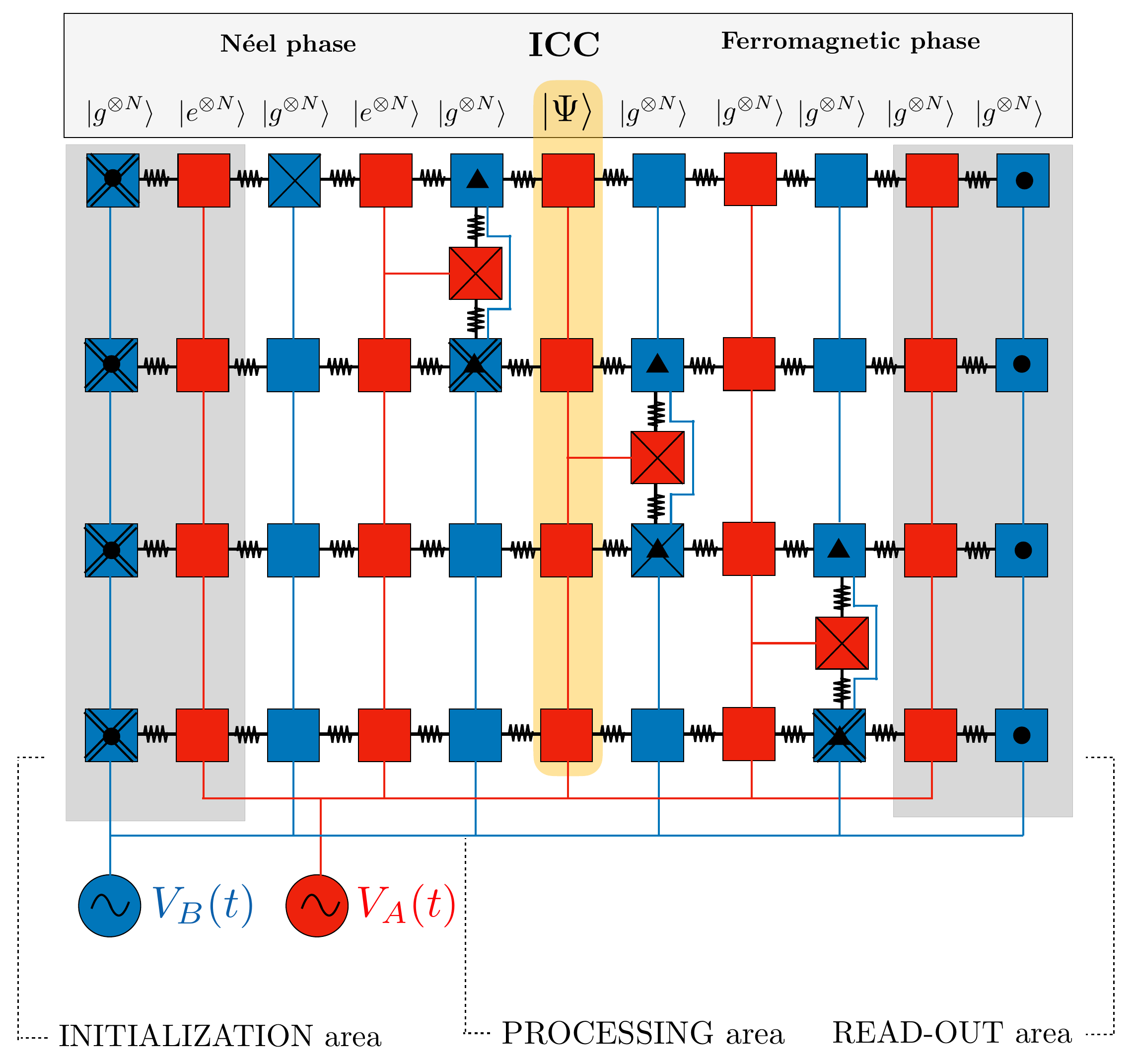}
\caption{Schematic description of a globally driven 2D ladder quantum computer with two species of qubits, an alternative version with respect to the three-species architecture described in Ref.~\cite{menta2024globally}. Two species, $A$, $B$ of superconducting qubits (red and blue squares, respectively) occupy the horizontal rows of a 2D ladder. Black springs and
 colored continuous lines represents respectively ZZ interactions among neighbor qubits and control lines. 
 Element with crosses and double crosses correspond to crossed and double crossed-qubits.  Filled circles and triangles on a square
emphasize that the corresponding site has coordination number 1 or 3, respectively (the other sites have all coordination number 2) -- see Eq.~(\ref{coordination}). Electrical wires depicted as continuous red and blue lines connect all elements within each group, facilitating global control through independent classical electrical pulses. During the computation, the logical information is encoded in the qubits of one of the columns of the processing area (highlighted in yellow in the figure): qubits on the left (right) hand side of such information carrier column  are in a ``N\'eel'' (``ferromagnetic'') $geg$ ($ggg$) phase. The illustration pertains to a $n=4$ computational qubit  quantum computer.
\label{fig:square}}
\end{figure}

The 2D ladder is now composed of two species, $\mathcal{S}=\{A,B\}$. Qubits on horizontal lines interact via a ZZ coupling and belong to the two species in an alternating way. On the other hand, each column is made of qubits of the same species, disconnected from each other. 
Each row of the processing area contains one $B$-type crossed or double-crossed element.
In some column made of $B$-type qubits, an $A$-type crossed qubit connects two rows of the ladder, as shown in Fig.~\ref{fig:square}. 
The quantum information is encoded vertically along a column, called the “information carrying column" (ICC), such that $N$ rows are needed to encode $N$ computational qubits. The ICC defines an interface: on one side of it the qubits are in the $\vert ...gegeg \rangle$ state (“paramagnetic"~\cite{para} or more properly, “N\'eel" phase) while on the other they are in the $\vert gggg... \rangle$ state (“ferromagnetic phase"). This asymmetry can be used to rigidly shift the interface along the ladder, maintaining the state of the ICC and the two phases intact. 

We emphasize that Corollary~\ref{corollary1} enables the replacement of the third species, $C$, in the ladder architecture of Ref.~\cite{menta2024globally} with $B$-type double-crossed elements, as depicted in Fig.~\ref{fig:square}. This result highlights that the ability to target individual qubits by introducing inhomogeneities into the processor allows for a reduction in the minimal number of species required to just two, while keeping the total number of physical qubits fixed.
To prove that this system is indeed a universal quantum computer, we need to prove that we are able to perform a universal set of quantum gates. The details on how the gates are performed can be found in Ref.~\cite{menta2024globally} with the extension of Eq.~\eqref{double-crossed}. 

Here we give an intuition of how the quantum processor works (see Fig.~\ref{fig:widefig}(a)). We start with single-qubit gates. Each line of the computer contains either a crossed or a double crossed qubit inside the processing area, as can be seen in Fig.~\ref{fig:square}. These qubits define specific positions where the single qubit gates are performed. We describe the guiding principle with a concrete example, referring to the ladder in Fig.~\ref{fig:square}. To perform an Hadamard gate on the second computational qubit of the ICC, we first bring it in the third column of the processing area. Now, the second computational qubit is located at the double-crossed qubit of the second line. Thanks to Corollary~\ref{corollary1}, we know that the state of this qubit can be controlled independently from the state of all the other computational qubits (since they are all located on regular qubits). Thus, thanks to this, one can engineer a pulse that does nothing on every computational qubit except for the second one which experiences a Hadamard gate. The same process can be done for every other computational qubit.

We now move on to two-qubit gates. These are mediated by $A$-type crossed qubits located between two lines. To entangle the third and fourth qubit of the ICC, we move it on the last column of the processing area. Once there, we apply a pulse that acts only on the crossed $A$-type qubits and induces a $2\pi$ rotation. This rotation is equivalent to a $-1$ phase factor on each crossed $A$-type qubit. For every one of these qubits except the one in the ICC, this factor is irrelevant. On the other hand, due to the blockade effect, the one in the ICC goes back to its initial state picking up a $-1$ factor only if its two neighbors, i.e., the third and fourth qubits, are both in the ground state. It is straightforward to see that this is sufficient to entangle the two computational qubits (the gate is a \texttt{CZ} gate) and thus provides the final ingredient to get universal quantum computation.

\subsection{Conveyor-belt with better scaling}

The second example of a globally controlled quantum computing scheme is the conveyor-belt architecture, initially proposed in Ref.~\cite{cioni2024conveyorbelt}. This architecture improves the scalability of the previously introduced 2D ladder schemes~\cite{menta2024globally, cesa2023universal}. Specifically, due to its one-dimensional closed-loop arrangement, which comprises two species of qubits, it requires $\mathcal{O}(n)$ physical qubits to encode $n$ computational qubits.
In this section, we introduce a two-species variant of the conveyor-belt architecture, which achieves even better scaling by reducing the required number of physical qubits by a factor of $\frac{1}{2}$.

\begin{figure}[t!]
\centering
\includegraphics[width=1.0\columnwidth]{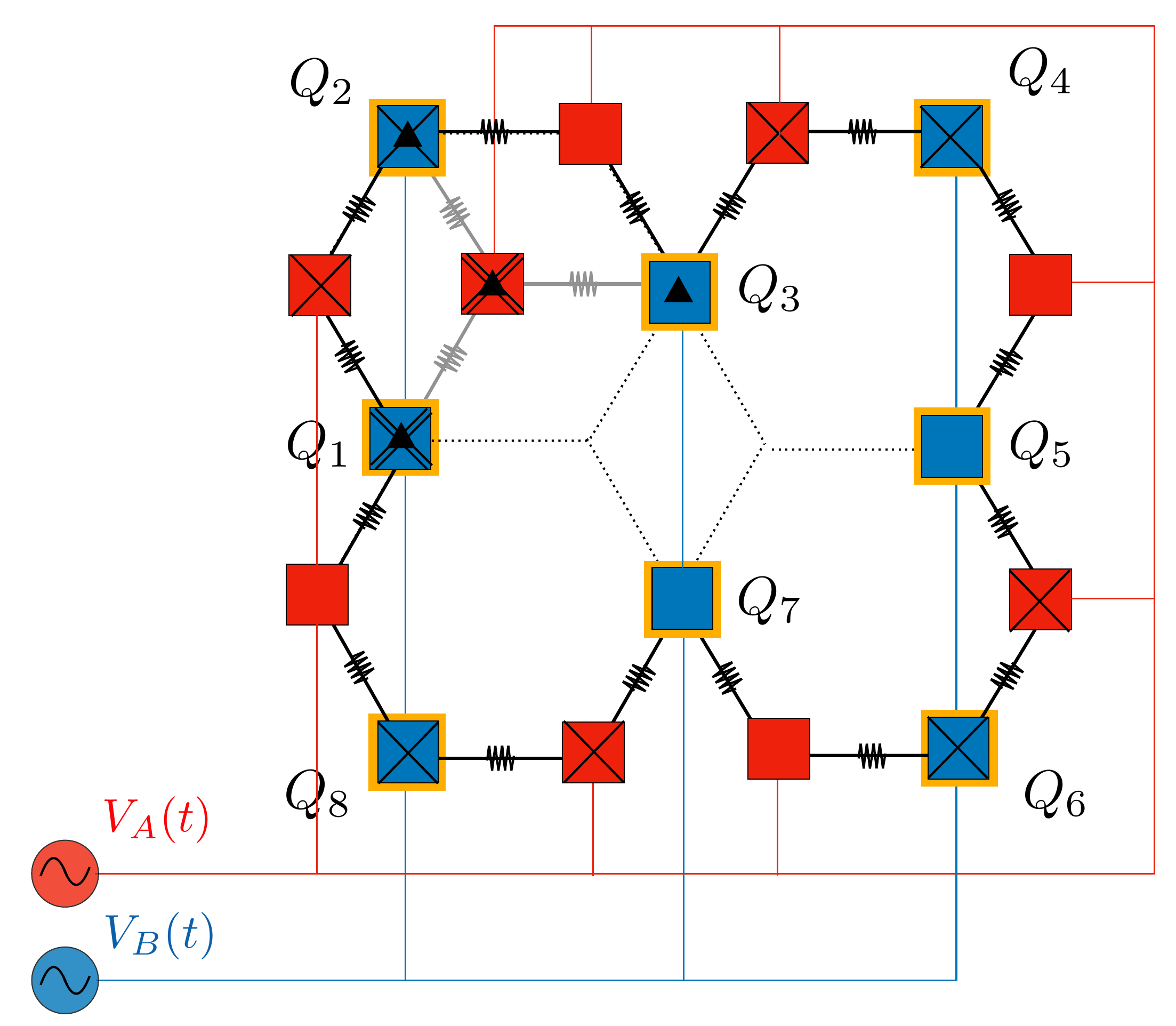}
\caption{
\label{fig:conveyorbelt} Schematic description of an alternative two-species variant to the conveyor-belt superconducting quantum computer introduced in  Ref.~\cite{cioni2024conveyorbelt}. Two types, $A$ and $B$ of superconducting qubits (red and blue squares, respectively) are separately driven by three classical sources $V_{A,B}(t)$  (red and blue continuous lines). They are coupled via a longitudinal ZZ coupling (black and grey springs). Black springs, crosses and double crosses,  filled circles and triangles are defined as in Fig.~\ref{fig:square}.
The $A$-type double-crossed qubit (red square inside  the loop) enables one-shot Toffoli gate (three-qubit gate) -- the corresponding 
interactions are depicted in gray. The $B$-type double crossed qubit performs universal single-qubit gates. 
The elements highlighted in yellow indicate the information carrying sites  $Q_1$, $Q_2$, $\cdots$, $Q_n$, separated by alternating $A$--type regular and crossed qubits.
The $Q_j$'s host the computational qubits through well-formed configurations.
The figure pertains to a $n=8$ qubit quantum computer.}
\end{figure}

The proposed architecture, which incorporates two species of qubits, $\mathcal{S} = \{A, B\}$, along with crossed and double-crossed elements, is illustrated in Fig.~\ref{fig:conveyorbelt}. Logical information, represented by $n$ computational qubits, is carried by $Q_1, \ldots, Q_n$ sites, is delocalized along the loop within the $B$-type elements. Regular and crossed $A$-type qubits, which are globally controlled and regularly spaced along the loop, enable the swapping dynamics of computational qubits. 
Here, we emphasize that, unlike Ref.~\cite{cioni2024conveyorbelt}, these swaps are implemented as two-qubit gates acting between neighboring computational qubits whose interaction is mediated via the single physical qubit between them. In particular, it is well known~\cite{Nielsen2010} that a $\texttt{SWAP}$ gate acting on two qubits, for instance 1 and 2, can be decomposed as $\texttt{CNOT}_{12} \texttt{CNOT}_{21} \texttt{CNOT}_{12}$, where the first and second subscripts denote the control and target qubits, respectively. 

As shown in Ref.~\cite{menta2024globally}, given a triple of qubits $BAB$ interacting via nearest-neighbor ZZ interactions, one can induce a native $\texttt{CZ}$ gate between the two external $B$-type qubits by applying a $2\pi$ rotation solely on the central mediator $A$-type qubit. Now, since $\texttt{CNOT}_{12} = \texttt{H}_2 \texttt{CZ}_{12} \texttt{H}_2$, a $\texttt{SWAP}$ between two computational $B$-type qubits can be implemented using the following protocol: \\

\begin{quantikz}
\lstick{$Q_{\rm e/o}$} & \swap{1} & \qw \\
\lstick{$Q_{\rm o/e}$} & \targX{} & \qw
\end{quantikz}
$\equiv$
\begin{quantikz}
\lstick{} & \ctrl{1} & \targ{}  & \ctrl{1} & \qw \\
\lstick{} & \targ{}  & \ctrl{-1} & \targ{}  & \qw
\end{quantikz}
$\equiv$
\begin{quantikz}
\lstick{} & \qw & \ctrl{1} & \gate{H} & \ctrl{1} & \gate{H} & \ctrl{1} & \qw \\
\lstick{} & \gate{H} & \ctrl{-1} & \gate{H} & \ctrl{-1} & \gate{H} & \ctrl{-1} & \gate{H}
\end{quantikz}
\\

Referring to Fig.~\ref{fig:conveyorbelt}, we define the qubit indexing as follows: (a) $Q_{\rm odd/\{1\}} \in B^{\rm r}$, (b) $Q_{\rm even} \in B^{\times}$ and (c) $Q_1 \in B^{\mathbb{X}}$ (used for computational single-qubit gates).

The conveyor-belt architecture we consider follows a periodic structure:
\begin{equation}
\underbrace{B^{\rm r}A^{\times}B^{\times}A^{\rm r}} \underbrace{B^{\rm r}A^{\times}B^{\times}A^{\rm r}} \ldots
\end{equation}
According to Corollary~\ref{corollary1}, our global drives $V_{A,B}(t)$ allow full independent control over all four elements of each block. Each periodic block can be decomposed into two overlapping interaction triples:
\begin{itemize}
\item[(T1)] $B^{\rm r}A^{\times}B^{\times}$, enabling a $\texttt{CZ}$ (and hence $\texttt{SWAP}$) between odd and even qubits.
\item[(T2)] $B^{\times}A^{\rm r}B^{\rm r}$, enabling a $\texttt{CZ}$ (and hence $\texttt{SWAP}$) between even and odd qubits.
\end{itemize}

Let us denote by $\hat{U}^{\rm swap}_{Q_i Q_j}$ the unitary operation performing a swap between qubits $Q_i$ and $Q_j$, mediated by the physical qubit in the corresponding triple. Each of these unitaries can be decomposed into the form of Eq.~\eqref{double-crossed}, with $\chi = A, B$.

To induce a movement of the logical qubits along the loop, we alternate the application of swap operations on the two types of triples. For a single time step $\ell$, we apply:
\begin{equation}
\hat{U}_\ell = \hat{U}^{\rm swap}_{Q_{\rm e}Q_{\rm o}} \hat{U}^{\rm swap}_{Q_{\rm o}Q_{\rm e}},
\end{equation}
where $\hat{U}^{\rm swap}_{Q_{\rm o}Q_{\rm e}}$ denotes the set of swaps applied to all disjoint triples of the form $B^{\rm r}A^{\times}B^{\times}$, and $\hat{U}^{\rm swap}_{Q_{\rm e}Q_{\rm o}}$ those on $B^{\times}A^{\rm r}B^{\rm r}$. After $T$ time steps, the total evolution is:
\begin{equation}
\hat{W}_T^{\circlearrowright_o, \circlearrowleft_e} = \prod_{\ell = 1}^T \left( \hat{U}^{\rm swap}_{Q_{\rm e}Q_{\rm o}} \hat{U}^{\rm swap}_{Q_{\rm o}Q_{\rm e}} \right),
\end{equation}
where the product is ordered left-to-right with increasing $\ell$. By increasing $T$, this composite unitary effectively realizes a cyclic permutation of the qubits. Odd-indexed qubits ($Q_{\rm odd}$) rotate in the clockwise direction, while even-indexed qubits ($Q_{\rm even}$) rotate counterclockwise. Importantly, by appropriately tuning $T$, any computational qubit can be moved to any desired logical position in the loop.

Let us make this more concrete. Let $|Q_1 Q_2 \dots Q_n\rangle$ be the initial configuration of $n$ computational qubits (with $n$ even, for geometric reasons). We define a transport operation $\hat{P}_{i \to j}$ that moves the state of the qubit $Q_i$ to the site $Q_j$ using a sequence of swap gates:
\begin{equation}
\hat{P}_{i \to j} = \prod_{(k, k+1) \in \mathcal{S}_{i \to j}} \hat{U}^{\rm swap}_{Q_k Q_{k+1}},
\end{equation}
where $\mathcal{S}_{i \to j}$ is the ordered list of adjacent pairs needed to transport $Q_i$ to $Q_j$ through nearest-neighbor swaps within its parity group (odd or even).

Since $\hat{W}_T^{\circlearrowright_o, \circlearrowleft_e}$ realizes a rigid cyclic shift among either odd- or even-indexed qubits, we can route any individual qubit to any desired logic site within its parity class. In particular, any computational qubit can be brought to the special $B^{\mathbb{X}}$ site for single-qubit operations, or to a Toffoli-capable triple for two-qubit gates.

However, it is important to note that this mechanism does not enable arbitrary permutations over all $n$ qubits. The conveyor-belt protocol implements only structured, parity-preserving rotations. Still, for the purpose of universal computation, this suffices: any qubit can be brought to a logic site where the desired gate is performed, and then returned to its original or a new location as needed. The architecture is thus fully logically addressable, albeit with sequential access.



\begin{figure*}[t!]
\centering
\includegraphics[width=1.0\textwidth]{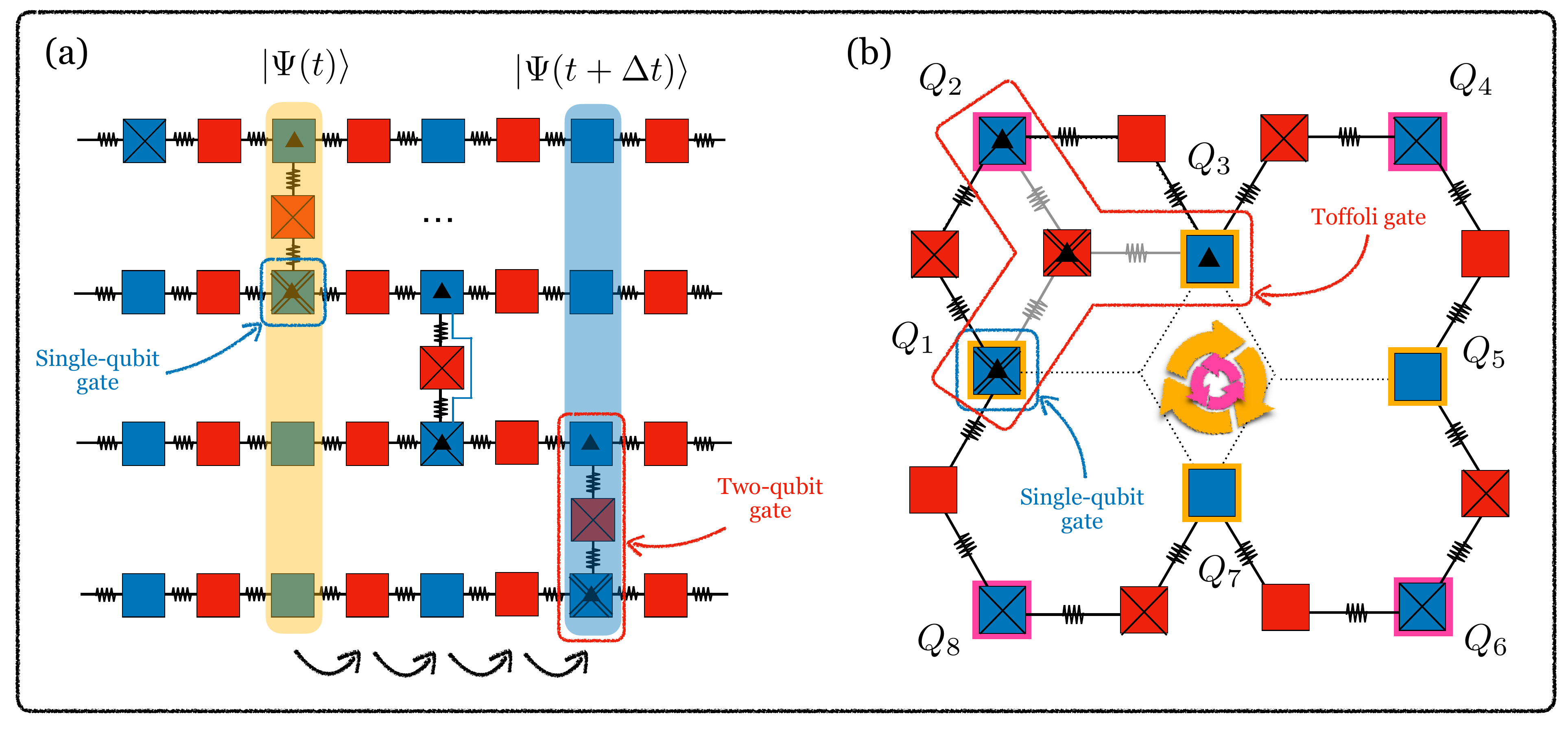}
\caption{Dynamics and quantum computation scheme of (a) 2D ladder architecture and (b) conveyor-belt architecture.}
\label{fig:widefig}
\end{figure*}

Despite the structural differences in implementation, the induced dynamics is functionally equivalent to that of Ref.~\cite{cioni2024conveyorbelt}. In summary, as shown in Fig.~\ref{fig:widefig}(b), global pulses induce rotational motion of the computational qubits: even-indexed qubits rotate counterclockwise, and odd-indexed qubits rotate clockwise along the loop. Naturally, the reverse direction can also be achieved by inverting the pulse sequence.

Let us now discuss how quantum computation is performed in this architecture. Referring to Fig.~\ref{fig:conveyorbelt}, the regular and crossed $B$-type elements are responsible for executing dynamical swap operations, which allow the transport of computational states along the chain. The double-crossed $B$-type qubit serves as the designated site for single-qubit gate operations, while the double-crossed $A$-type qubit enables the execution of one-shot Toffoli gates. As demonstrated in Ref.~\cite{cioni2024conveyorbelt}, this set of capabilities, e.g. the ability to perform single-qubit gates and Toffoli gates, is sufficient to achieve computational universality.

For instance, suppose that we wish to apply a Hadamard gate to the computational qubit currently encoded in $Q_5$. By applying global pulses that activate swaps across alternating layers of qubits, the state stored in $Q_5$ can be dynamically routed to $Q_1$, which hosts the unique double-crossed $B$-type qubit. Once the computational qubit arrives at this site, Corollary~\ref{corollary1} ensures that we can address it individually and implement any desired single-qubit unitary, including the Hadamard gate.

Similarly, consider the case where we want to entangle two computational qubits, say those carried by $Q_6$ and $Q_7$. Using a sequence of swap operations, we can transport their states, $|\psi_6\rangle$ and $|\psi_7\rangle$, to the neighboring sites $Q_2$ and $Q_3$, respectively. These two qubits then form part of a local triple with the crossed $A$-type qubit situated in between. By sequentially applying the global pulses that activate the relevant species and leveraging Corollary~\ref{corollary1}, a native three-body interaction can be used to implement a Toffoli gate. From this, a $\texttt{CZ}$ gate can be efficiently synthesized following standard circuit identities~\cite{Nielsen2010}.

This example illustrates how quantum algorithms can be systematically executed within the conveyor-belt quantum processor. By combining dynamical routing via global swaps with localized universal gate sets at designated sites, the architecture supports full logical addressability and universal quantum computation.

\subsection{Summary}
We conclude the main section with a brief comparison of the two architectures -- the 2D ladder and the conveyor-belt -- in terms of resource requirements. To provide a clearer understanding, we present Table~\ref{tab1} below, which highlights the differences between these architectures and their proposed variants.

\begin{table}[!htp]
\centering
\begin{tabular}{c|cccc}
& Species & Physical qubits & Crossed & Double-crossed \\
\hline
Ref.~\cite{menta2024globally} & 3  & $2n^2 + 4n - 1$  & \cmark & \xmark  \\
Ref.~\cite{cioni2024conveyorbelt} & 2 & $4n + 1$ & \cmark & \xmark \\
Fig.~\ref{fig:square} & \textcolor{red}{2} & $2n^2 + 4n - 1$ & \cmark & \textcolor{red}{\cmark} \\
Fig.~\ref{fig:conveyorbelt} & 2 & \textcolor{red}{$2n + 1$} & \cmark & \textcolor{red}{\cmark} \\
\end{tabular}
\caption{Summary of the different configurations of species, physical qubits, and crossing elements for the models. The “crossed" column indicates whether crossed elements are present (\cmark) or not (\xmark), and the “double-crossed" column shows if double-crossed elements are included. $n$ is the number of computational qubits. In red the novelty with respect to previous implementations.}
\label{tab1}
\end{table}

\section{Conclusion}
\label{sec:conclusion}

In this work, we analyzed the core requirements for constructing globally controlled quantum processors, focusing on architectures compatible with superconducting qubits and native ZZ interactions. While global control offers a promising solution to the wiring bottleneck in large-scale quantum systems, it inherently limits addressability, making the implementation of logical operations nontrivial. To overcome this, we identified the necessity of introducing anisotropic elements -- such as crossed and double-crossed qubits -- into the qubit array. These elements serve as symmetry-breaking sites where global drives can induce localized effects, thus enabling targeted gate operations despite the uniform nature of the control fields.
We formalized this mechanism in Theorem~\ref{theorem1}, which establishes conditions under which a global pulse can effectively act on a selected subsystem. Corollary~\ref{corollary1} further demonstrates how this principle extends to chains of interacting qubits, ensuring simultaneous and independent controllability of key architectural elements.

These theoretical results were then applied to two globally driven architectures. In both cases, logical gate execution is achieved by transporting computational qubits through swap operations toward special logic sites. The second example, the conveyor-belt architecture, implements this logic in a scalable and symmetric layout. It exploits alternating swap layers and structural asymmetry to move qubits efficiently and perform universal gate sets at designated locations. Notably, it achieves a scaling in the number of required physical qubits that, to the best of our knowledge, is the most efficient ever realized in a global control setting. In a realistic scenario, inhomogeneities in the coupling strengths and qubit frequencies affect the achievable fidelities. It has been recently demonstrated that such disorder can be mitigated by employing optimal control pulses~\cite{Aiudi2025, SethQAOA, Hu2025}.

Although the ladder and conveyor-belt architectures feature an intrinsically one-dimensional layout, their reconfigurable and dynamical character enables effective long-range interactions that emulate higher-dimensional connectivity over time. This property allows the integration of error-correcting schemes compatible with 1D connectivity—such as concatenated, subsystem, or LDPC-based codes~\cite{LDPC, Panteleev2021}—and supports the implementation of fault-tolerant protocols through temporal multiplexing~\cite{Litinski2019}. The search of tailored quantum error correcting codes for these architectures is part of current research in the field of globally-controlled quantum computing~\cite{Bririd_2004, Kay_2005, Kay_2007, Fitzsimons_2007, Fitzsimons_2008}.

Our results clarify the fundamental trade-offs in globally controlled quantum computing and outline a concrete framework for designing scalable, programmable quantum processors with reduced wiring complexity. This architecture bridges conceptual simplicity with practical feasibility and may serve as a blueprint for future hybrid systems that blend global control with minimal local addressability.

\acknowledgments
This work has been funded by the European Union - NextGenerationEU, Mission 4, Component 2, under the Italian Ministry of University and Research (MUR) Extended Partnership PE00000023 National Quantum Science and Technology Institute -- NQSTI -- CUP J13C22000680006. 
M.P. and V.G. are co-founders and shareholders of Planckian.
At the time of their contributions, authors affiliated with Planckian are either employees at Planckian or PhD students collaborating with Planckian.


\begin{thebibliography}{99}
\bibitem{shift}
This shift works when the ZZ coupling is homogeneous, but it can be generalized to $\zeta_{ij}$ if the ZZ interactions are non-homogeneous.
\end{thebibliography}
\end{document}